\documentclass[aps,pre,twocolumn,letter]{revtex4-1}
\usepackage{graphicx}
\usepackage{float}
\usepackage{amsmath}
\usepackage{color}
\usepackage{xr}
\begin{document}
\title{Hidden symmetries in jammed systems}
\date{\today}
\author{Peter K. Morse and Eric I. Corwin}
\affiliation{Department of Physics and Materials Science Institute, University of Oregon, Eugene, Oregon 97403, USA.}

\begin{abstract}
There are deep, but hidden, geometric structures within jammed systems, associated with hidden symmetries.  These can be revealed by repeated transformations under which these structures lead to fixed points.  These geometric structures can be found in the Voronoi tesselation of space defined by the packing.  In this paper we examine two iterative processes: maximum inscribed sphere (MIS) inversion and a real-space coarsening scheme.  Under repeated iterations of the MIS inversion process we find invariant systems in which every particle is equal to the maximum inscribed sphere within its Voronoi cell.  Using a real-space coarsening scheme we reveal behavior in geometric order parameters which is length-scale invariant.
\end{abstract}

\maketitle

\section{Background}

The jamming of athermal spheres has been called the epitome of disorder \cite{ohern_jamming_2003}, but this doesn't mean that it is devoid of all symmetries. The structure found within jammed systems distinguishes them from Poisson point processes, for which there is no structure or correlation. This structure is reflected in the non-trivial behavior of correlations (such as the pair-correlation function) \cite{bernal_radial_1962, donev_pair_2005}, geometry \cite{morse_geometric_2014, morse_geometric_2016}, contact number distributions \cite{maxwell_calculation_1864}, hyperuniformity \cite{donev_unexpected_2005, dreyfus_diagnosing_2015}, and volume distributions \cite{clusel_granocentric_2009, maiti_free_2014}.  Recent theoretical work has demonstrated the existence, and breaking of an abstract replica symmetry at the jamming and glass transitions, namely liquid systems have replica symmetry and glasses or jammed systems break that symmetry.  \cite{parisi_mean-field_2010, charbonneau_fractal_2014}.  In this work we search for evidence of hidden symmetries in the spatial and geometric structure of systems below, at, and above jamming.  We employ the general scheme of repeated transformations in the hope that they will lead to fixed point systems reflective of the underlying symmetries.

We have previously shown that a number of geometric properties of the Voronoi tesselation carry signatures of the jamming transition, including the number of neighbors, surface area, volume, aspect ratio, and maximum inscribed spheres (MIS). There is an obvious symmetry that jumps out precisely at the jamming transition.  Because each particle in a jammed system is in contact with several of its neighbors it must kiss the boundaries of its Voronoi cell.  Therefore, the maximum inscribed sphere of a Voronoi cell must be equal to the particle. This suggests a transformation of repeatedly replacing every particle with the maximum inscribed sphere of it's Voronoi cell.  Under such a transformation, jamming will necessarily be a fixed point.

The renormalization group has been very successful in understanding phase transitions.  At present, jamming lacks an appropriate field with which To construct a renormalized field theory.  Nevertheless, we barrel ahead with a brute force real space coarsening and rescaling scheme in the hopes of pointing the way to the appropriate field theory.  By repeated iteration of this coarsening we test a range of geometric order parameters to see if they mimic the behavior of a proper renormalized field.

\begin{figure}[h]
\includegraphics[width=1\linewidth]{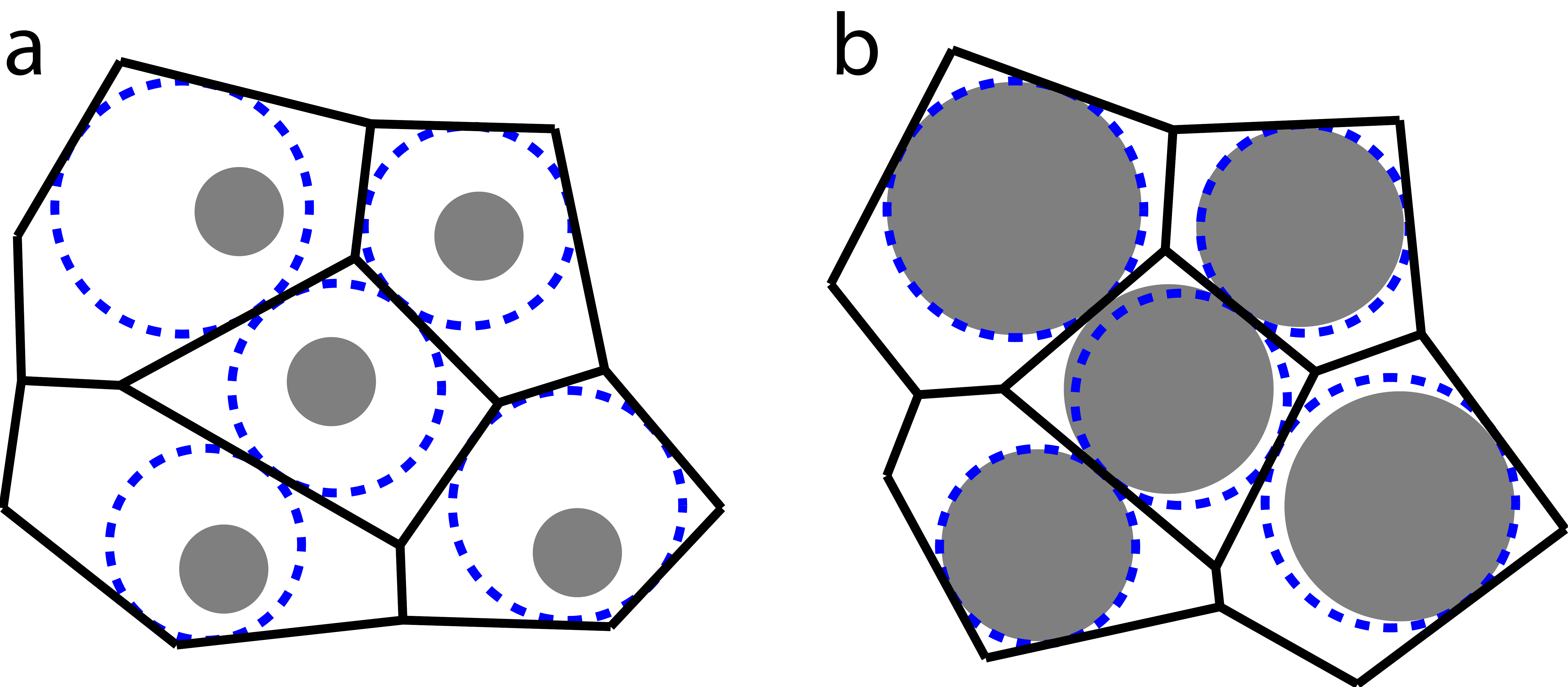}
\caption{Diagram showing the maximum inscribed sphere (MIS) inversion. (a) An initial monodisperse packing is shown with particles colored gray, their radical Voronoi cells in black, and the MIS shown as a dashed blue line. (b) The first step of the MIS inversion with the same color scheme. Note that the MIS from (a) is the new particle, the Voronoi cell has changed, and the system is now polydisperse.}
\label{fig:MISIllustration}
\end{figure}

\section{Methods}

\subsection{Packing Simulations}

In order to generate the initial packings for our MIS Inversion and coarse graining, we use an infinite temperature quench protocol \cite{ohern_random_2002} as described in reference \cite{morse_geometric_2014}. The process begins with uniformly randomly distributed athermal fricitonless particles with monodisperse or polydisperse radii chosen to achieve a desired packing fraction.  The packing fractions for monodisperse packings are chosen within the range $\phi = 0$ to $\phi = 2\phi_J$ (where $\phi_J$ is the packing fraction at jamming) to obtain the full spectrum of behavior, both well below and well above jamming. Polydisperse packings are generated with densities from $\phi=0$ to significantly above $\phi_J$ but not all the way to $2\phi_J$.  We choose a harmonic contact potential between particles and then allow the system to relax to its local potential energy minimum (also called its inherent structure) via the conjugate gradient algorithm \cite{hestenes_methods_1952} for systems below $\phi_J$ or the fast inertial relaxation engine (FIRE) algorithm \cite{bitzek_structural_2006} for systems above $\phi_J$. For studies of the MIS Inversion we use monodisperse packings of 8192 particles and polydisperse packings of 16384 particles.  Coarse graining requires higher numbers of particles, and so we use monodisperse systems that contain 65536 particles.  All data is monodisperse unless otherwise stated.

For polydisperse packings, we choose radii from a log-normal distribution. We characterize a polydisperse packing by the standard deviation of the radii divided by the mean. For the MIS inversion, we report on initial polydispersities of 0, 0.05, 0.1, 0.15, and 0.2.

\subsection{Geometric Cell Properties}

The maximum inscribed sphere (MIS) of a Voronoi cell is the largest sphere that is fully contained within the boundaries of the cell. It is calculated using linear programming techniques \cite{boyd_convex_2004}.

Detailed descriptions of how we calculate the number of neighbors, the surface area, the volume, and the aspect ratio can be found in our previous work \cite{morse_geometric_2014, morse_geometric_2016}. Two particles are considered neighbors if their Voronoi cells share a facet, so to find the number of neighbors, we calculate the number of facets via a method developed by Boissonnat \cite{boissonnat_convex_2005}. To find the surface area and the volume, we calculate the veritices of each cell and take a delaunay triangulation to split each facet into simplicies. From these simplices and an interior point of the cell, it is simple to calculate the surface area and volume associated with each cell. 

The aspect ratio that we use for convex cells is defined by the maximum distance between any pair of vertices contained in a cell divided by the minimum of the maximum distance between each vertex and every facet.  Thus, the aspect ratio can be seen as the ratio between the longest and shortest spanning lengths of a cell.

\subsection{Maximum Inscribed Sphere Inversion}

The MIS inversion process starts with a sphere packing at any density.  From this packing we calculate the radical Voronoi tesselation (also sometimes called the Laguerre tesselation).  We choose to use the radical Voronoi tesselation as opposed to the additively-weighted Voronoi tesselation (or indeed any other tesselation) to ensure that the Voronoi cells are always convex regardless of polydispersity \cite{voronoi_nouvelles_1908}.  From this tesselation we calculate the MIS for each cell.  Finally, this new set of spheres is treated as a new sphere packing upon which we can iterate this procedure.  We note that each MIS is in general uniquely determined, with the exception of the pathological case in which there is a high degree of symmetry in the underlying Voronoi cell (for example, if the cell is a rectangular solid). This pathological case will not happen in a disordered system, and so we do not explicitly account for it.  However, even if such cases were present they would simply manifest as a degenerate set of spheres, of which our analysis would choose one.

\begin{figure}[h]
\includegraphics[width=1\linewidth]{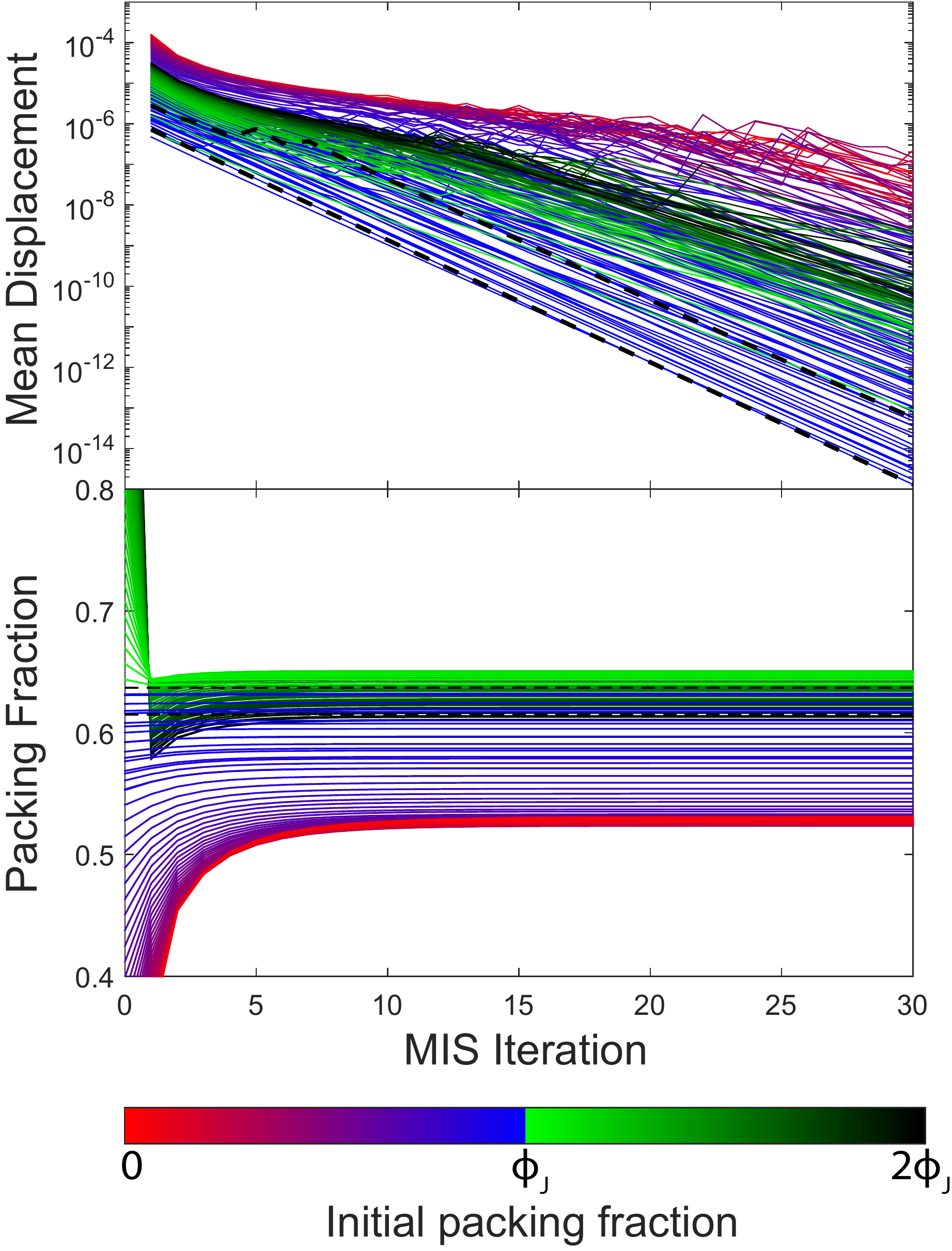}
\caption{Convergence of the MIS inversion.  a) Log-linear plot of the mean displacement per particle from the previous MIS iteration. Each line represents a system at a different initial packing fraction, with color scale indicating initial packing fraction as shown beneath the plots.  The lower dashed line is drawn at $\phi_J$ and the upper dashed line is drawn at $\phi^*$, the lowest packing fraction for which initial packings are fixed points under the MIS inversion. b) Packing fraction at each step of the MIS inversion, using the same color scheme as a.  The upper dashed line is drawn here at $\phi_J$ and the lower dashed line is at $\phi^*$, which is flipped from a.}
\label{fig:MISConvergence}
\end{figure}

We find that for all systems this process converges to a fixed point packing, dependent on the initial input.  In practice, we find that repeating this process $30$ times is sufficient to find fixed points in which each particle deviates by less than one part in $10^6$ from it's former position, as shown in figure \ref{fig:MISConvergence}a.

Even though the MIS inversion constructs packings that are guaranteed to have no overlaps we can still define a coordination number $z$ for each particle by choosing an appropriate cutoff distance.  We pick this distance to fall immediately after the first peak in the pair-correlation function.

\subsection{Real-Space Coarsening}

\begin{figure}[h]
\includegraphics[width=1\linewidth]{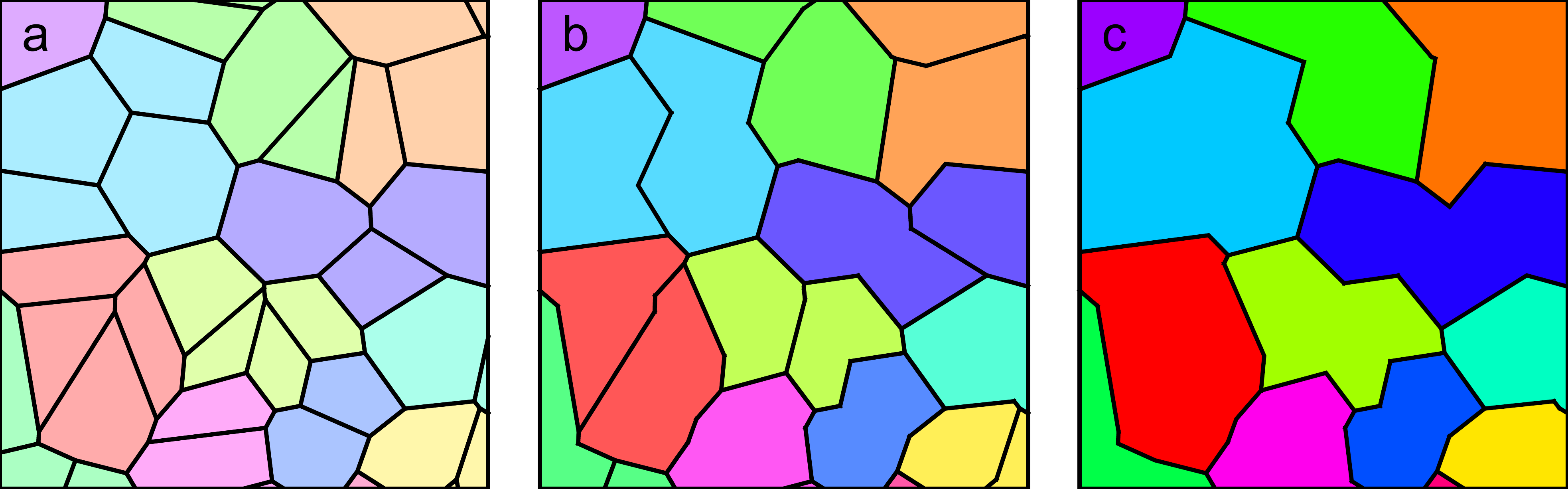}
\caption{The coarse graining procedure pairs each cell in with the unpaired cell that shares the highest surface area. a) The original Voronoi diagram, b) the system after one step of the coarsening procedure, c) the system after two steps of coarsening. Colors denote the membership in cells after two rounds of coarsening.}
\label{fig:renormIllustration}
\end{figure}

Given a sphere packing at any density, we compute the Voronoi tesselation and the geometric cell properties, defined above.  Because the systems under study here are monodisperse it doesn't matter which Voronoi tesselation we choose, as they are all degenerate. Our real space coarsening scheme creates a coarser grained packing by joining pairs of neighboring cells.  To do so, we choose a random ordering of cells and sequentially pair each cell in our list with the the unpaired neighboring cell that shares the largest interfacial surface area. Both cells are then removed from the list, and the process continues until nearly all cells are paired. Because of the random nature of this process there will always be a few singleton cells remaining which we join with the previously paired cell with which it shares the highest interfacial surface area (with the caveat that no composite cell can be made up of more than three cells). These triplet cells constitute less than $3.5\%$ of the total cells in $d=3$, $1.7\%$ in $d=4$, and $0.7\%$ in $d=5$.  The set of paired cells covers all space and thus produces a new tesselation.  Note that this new tesselation is \textit{not} a Voronoi tesselation and is not even convex. However, our pairing scheme is designed to create cells that are compact and nearly convex upon high iterations.  We continue this process iteratively to create a further new tesselation with approximately half as many cells as the previous one.  Thus the number of iterations possible in this scheme is limited by the number of particles in the initial packing. In $d = 3$, we average this process over 100 different orderings. In $d=4$, we use one ordering due to the increased computation time.

If we label two neighboring cells $A$ and $B$ we can call the combined cell $A \cup B$.  The neighbors of $A \cup B$ are the union of the set of neighbors of $A$ and $B$ (not counting each other), the volume is simply the sum of the constituent volumes, and the surface area is the sum of the constituent surface areas minus twice the interfacial surface area. Because the new cells are not convex and our aspect ratio only applies to convex cells, we take the convex hull of the vertices of all the constituent cells and we calculate the aspect ratio as defined above. This preserves the definition that the aspect ratio is the longest one dimensional distance divided by the shortest one dimensional distance, but these two distances are no longer constrained to be fully contained within the cell. 

When a combined cell is made of three constituent cells $A$, $B$, and $C$, this process is done for $A \cup B$ and then $(A \cup B) \cup C$.

%
%\begin{equation}
%N_{A \cup B} = N_A \cup N_B.
%\end{equation}
%

%
%\begin{equation}
%V_{A \cup B} = V_A + V_B.
%\end{equation}
%

%\begin{equation}
%S_{A \cup B} = S_A + S_B - 2S_{A \cap B}.
%\end{equation}
%

\begin{figure}[h]
\includegraphics[width=1\linewidth]{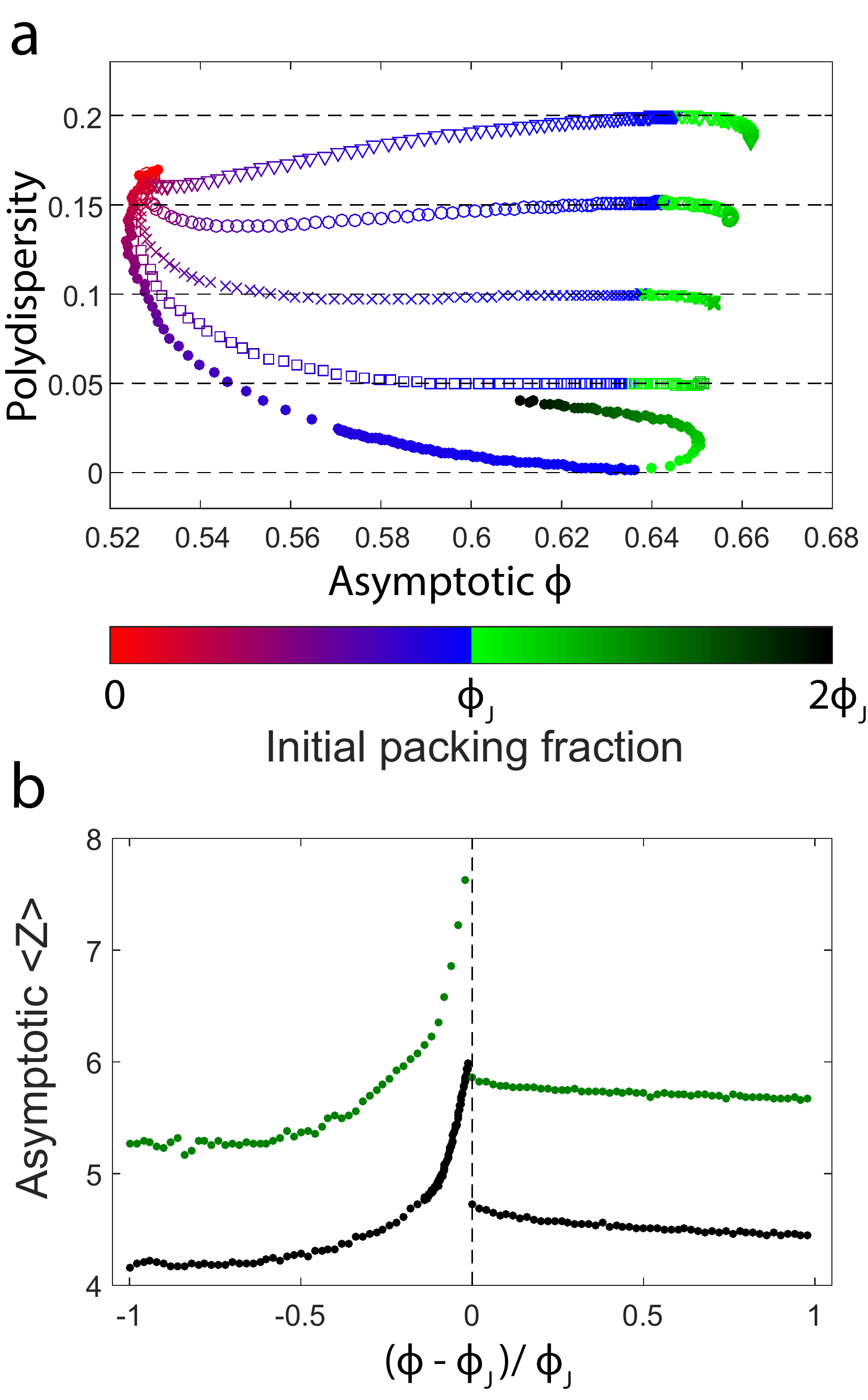}
\caption{Asymptotic properties of the MIS inversion.  a) Fixed points of the MIS Inversion are represented by their polydispersity and their final packing fraction. Initial packings start at polydispersities of 0 (closed circles), 0.05 (squares), 0.1 (x's), 0.15 (open circles), and 0.2 (triangles). Points that remain fixed through the entire process lie on the dotted lines associated with their starting polydispersity. Initial packing fraction is coded into the color. b) Asymptotic mean number of contacts $\langle Z \rangle$ as a function of initial packing fraction for $d=3$ (black), and $d=4$ (green).}
\label{fig:MISPolyVsPhi}
\end{figure}

\section{Discussion}
\subsection{Maximum Inscribed Sphere}

Under repeated action of the MIS inversion, we find that every initial packing quickly reaches a fixed point depending only on the initial packing fraction and initial polydispersity. Figure \ref{fig:MISPolyVsPhi}a plots the asymptotic polydispersity against the asymptotic packing fraction.  Each curve shown corresponds to a different starting polydispersity, ranging from 0 to 0.2, in steps of 0.05, with varied initial packing fraction from $\phi=0$ to $\phi=2\phi_J$ represented by the color scale.  These fixed points form a continuous line in this phase space, demonstrating that packings with similar initial properties to one another transform into fixed point packings with similar asymptotic properties.

The presence of rattlers (particles which are unconstrained, even in a jammed system) induces a small drift between the initial packings and fixed point packings, even at $\phi_J$.  If we were to apply this same transformation in higher dimensions where the presence of rattlers is greatly diminished \cite{charbonneau_universal_2012}, we expect that the points would correspondingly drift even less. 

By design, jammed packings at $\phi_J$ are a fixed point under this repeated transformation.  We find that systems extending a significant distance below jamming are also fixed points.   Surprisingly, this extends down to a density consistent with our previously discovered $\phi^*$ \cite{morse_geometric_2014}.  This is the point at which the distribution of MIS radii first becomes a delta function. The values used for $\phi^*$ are 0.615 and 0.427 in $d=3$ and $d=4$ respectively. However, we find that the mean coordination number in dimension d jumps from $d+1$ to $2d$ not at $\phi^*$, but at $\phi_J$, which simply indicates that the particles in this range are not jammed. The fact that systems between $\phi^*$ and $\phi_J$ are fixed points of this transformation points conclusively to a symmetry that exists only within this range: all non-rattler particles are exactly equal to the MIS of their Voronoi cells.

Systems that begin well above jamming get folded back to systems below jamming under this transformation. This is because the MIS of a Voronoi cell for an overjammed particle is always smaller than the particle itself due to the fact that overjammed particles are not fully contained within their Voronoi cells \cite{morse_geometric_2014}. These systems always appear to be folded onto the range between $\phi^*$ and $\phi_J$.

We observe a cluster of data points near an asymptotic $\phi = 0.53$, with no fixed points falling below this density.  This packing fraction is strikingly similar to those obtained in frictional random loose packing (RLP) experiments and simulations \cite{onoda_random_1990, song_phase_2008}.  Further, these fixed points all have a mean coordination number of $\langle Z \rangle \simeq 4$ in $d=3$ (shown in figure \ref{fig:MISPolyVsPhi}b), as required for RLP.  By design, each MIS particle must have a minimum of $Z = d+1$ non-cohemispheric contacts, but this only sets a lower bound.  Thus, these fixed point packings appear to be a frictionless analog of RLP.  In this way, MIS inversion offers an interesting, non-physical mechanism for generating polydisperse RLP packings.  It would be interesting to compare the same process in higher dimensions to high dimensional simulations of RLP if they were to become available.

\subsection{Real-Space Coarsening}

\begin{figure*}[h]
\includegraphics[width=1\linewidth]{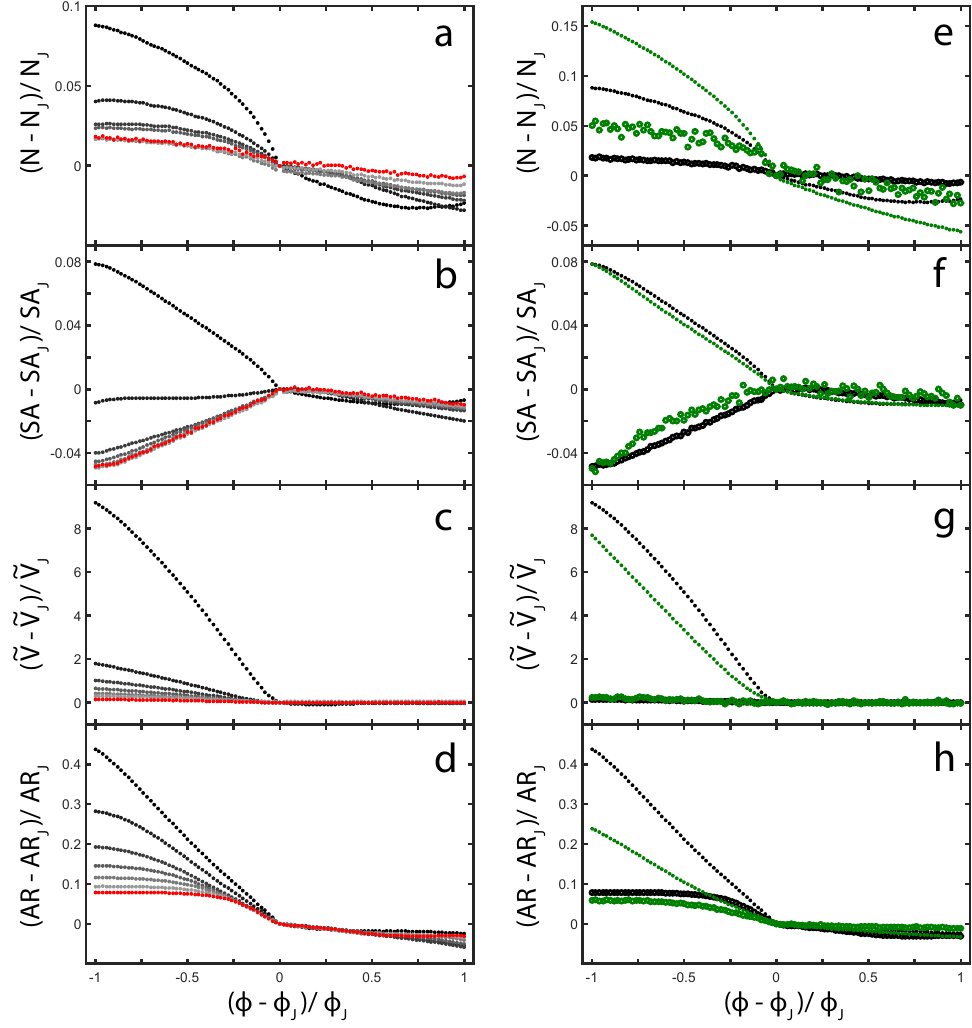}
\caption{Under the coarse graining scheme, the behavior of each parameter of the cells reaches a curve of fixed points as a function of packing fraction. On the left, we have plotted the initial parameter (black) and the first 6 coarse grainings (each corresponding to a lighter shade of gray, with the 6th in red) in three dimensions for the mean number of neighbors (a), mean surface area (b), standard deviation over the mean of volume (c) and aspect ratio (d). In each, we scale by subtracting off and dividing by the order parameter at $\phi_J$ in that iteration, such that the jamming point is always at the origin. On the right (e-h), we have plotted only the original (closed circles) and the 6th iteration (open circles) for $d = 3$ (black) and $d = 4$ (green).}
\label{fig:renormFirstAndLast}
\end{figure*}

For all of the parameters studied, our systems reach fixed points under appropriate rescaling after only a few iterations of the coarsening procedure.  However, the shape of the curve of fixed points appears to be strongly dependent on the parameters in question. Figure \ref{fig:renormFirstAndLast}a-d shows the iterative application of the coarse graining to a range of systems between $\phi=0$ and $\phi=2\phi_J$ in $d=3$ for the mean number of neighbors (a), the mean surface area (b), the standard deviation over the mean of the volume (c), and the mean aspect ratio (d). Figure \ref{fig:renormFirstAndLast}e-h shows just the original order parameter and the fixed point after 6 iterations of the coarsening for $d=3$ and $d=4$, showing similar behavior across dimension. 

The mean number of neighbors (Figure \ref{fig:renormFirstAndLast} a,e) retains its shape most closely, but is highly susceptible to noise, as each cell can only have an integer number of neighbors. The volume graph (Figure \ref{fig:renormFirstAndLast} c,g) appears to be even more susceptible to noise, as the standard deviation over the mean goes to zero upon even modest iterations of the process. It is the surface area and the aspect ratio which show the strongest and most interesting behavior. The surface area (Figure \ref{fig:renormFirstAndLast} b,f) appears to come into the transition linearly at first, but after the transformation, the trend reverses itself and quickly reaches a fixed point that is almost a mirror of the original. Interestingly, each packing fraction retains perfect memory of where it began before the iterations. The same statement about memory could possibly be said for the neighbors if the graphs weren't so susceptible to noise.

\begin{figure}[h]
\includegraphics[width=1\linewidth]{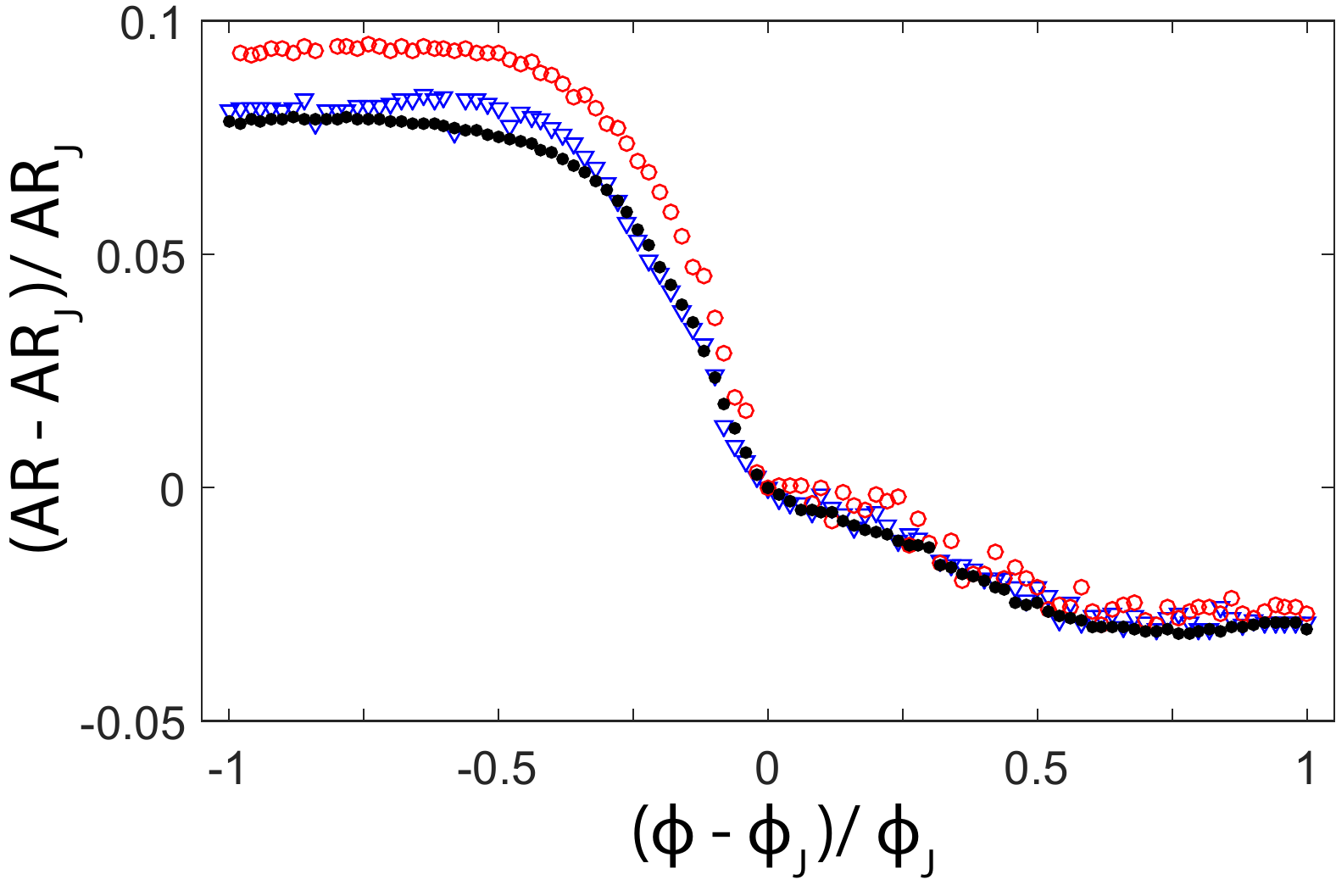}
\caption{Coarse grained aspect ratio cross over shows no evidence of finite size effects. We have plotted 3 system sizes in $d = 3$ at the 6th iteration of the coarsening (as done in figure \ref{fig:renormFirstAndLast}h), with system sizes of 4096 (red open circles), 8192 (blue triangles), and 65536 (black closed circles).}
\label{fig:renormFiniteSize}
\end{figure}

The aspect ratio (Figure \ref{fig:renormFirstAndLast} d,g) gives us the strongest signature of the jamming transition under repeated iterations of the coarse graining. We observe that at high iterations of the process we get one value of the aspect ratio far below jamming and one value above jamming. There is a cross-over region which may be related to the pre-jamming phase transition that occurs at $\phi^*$, as the halfway point seems to coincide with the numerical values used for $\phi^*$. By examining this behavior as a function of system size in figure \ref{fig:renormFiniteSize} we can conclude that this extended cross-over regime is not merely the finite size blunting of a step function.  Indeed, there is very little change at all between systems of 4096 particles and those of 65536.

\section{Conclusion}

Each of these transforms provide a new perspective towards understanding jammed systems. Through the MIS inversion, we have clarified the pre-jamming phase transition at $\phi^*$. Systems between $\phi^*$ and $\phi_J$ are fixed points of the inversion, meaning that they exhibit a symmetry wherein the MIS is equal to the particle itself. This symmetry is interesting in that it relates the real particles to the Voronoi network, which is a purely geometric construct. We posit that the transition at $\phi^*$ can only be seen when observing an order parameter that is influenced both by the mechanical network of contacts as well as the geometric Voronoi cell.  Thus, it is perhaps not surprising that this pre-jamming transition is absent in purely mechanical or purely geometric order parameters. It is interesting to note that the value of $\phi^*$ is very close to the value of the ideal glass transition in thermal hard spheres \cite{parisi_ideal_2005}. We are actively pursuing this line of research.

The coarse graining reveals a different symmetry which could be useful in constructing a renormalizable field theory for jamming. Both the surface area and the aspect ratio present interesting features under coarse graining which preserve knowledge of the underlying packing while other features are washed out.  Somehow, the surface area retains perfect memory of the underlying packing.  Aspect ratio is particularly intriguing as it separates systems into two clearly different classes, above and below the jamming transition.  This is reminiscent of the renormalization group's delineation between relevant and irrelevant graphs. If this can be translated into a rigorous renormalization group analysis, it would imply that either the surface area, the aspect ratio, or both reveal symmetries that fundamentally differentiate a jammed packing from an unjammed one.

%
%\begin{table}
%
%\begin{tabular}{c|c|c|c|c}
%d & 3 & 4 & 5 & 6\\
%\hline
%$\phi_{LR}$ & 0.5832 & 0.3948 & 0.2581 & 0.163\\
%\hline
%$\phi_J$ & 0.6437 & 0.4562 & 0.3079 & 0.1999\\
%\end{tabular}
%\caption{Values of jamming point and local rigidity onset point}
%\label{tab:N0}
%
%\end{table}

\begin{acknowledgments}
This work was supported by the NSF under Career Award DMR-1255370. The ACISS supercomputer is supported under a Major Research Instrumentation grant, Office of Cyber Infrastructure, OCI-0960354.
\end{acknowledgments}

\bibliography{JammingFixedPoints}

\begin{thebibliography}{22}%
\makeatletter
\providecommand \@ifxundefined [1]{%
 \@ifx{#1\undefined}
}%
\providecommand \@ifnum [1]{%
 \ifnum #1\expandafter \@firstoftwo
 \else \expandafter \@secondoftwo
 \fi
}%
\providecommand \@ifx [1]{%
 \ifx #1\expandafter \@firstoftwo
 \else \expandafter \@secondoftwo
 \fi
}%
\providecommand \natexlab [1]{#1}%
\providecommand \enquote  [1]{``#1''}%
\providecommand \bibnamefont  [1]{#1}%
\providecommand \bibfnamefont [1]{#1}%
\providecommand \citenamefont [1]{#1}%
\providecommand \href@noop [0]{\@secondoftwo}%
\providecommand \href [0]{\begingroup \@sanitize@url \@href}%
\providecommand \@href[1]{\@@startlink{#1}\@@href}%
\providecommand \@@href[1]{\endgroup#1\@@endlink}%
\providecommand \@sanitize@url [0]{\catcode `\\12\catcode `\$12\catcode
  `\&12\catcode `\#12\catcode `\^12\catcode `\_12\catcode `\%12\relax}%
\providecommand \@@startlink[1]{}%
\providecommand \@@endlink[0]{}%
\providecommand \url  [0]{\begingroup\@sanitize@url \@url }%
\providecommand \@url [1]{\endgroup\@href {#1}{\urlprefix }}%
\providecommand \urlprefix  [0]{URL }%
\providecommand \Eprint [0]{\href }%
\providecommand \doibase [0]{http://dx.doi.org/}%
\providecommand \selectlanguage [0]{\@gobble}%
\providecommand \bibinfo  [0]{\@secondoftwo}%
\providecommand \bibfield  [0]{\@secondoftwo}%
\providecommand \translation [1]{[#1]}%
\providecommand \BibitemOpen [0]{}%
\providecommand \bibitemStop [0]{}%
\providecommand \bibitemNoStop [0]{.\EOS\space}%
\providecommand \EOS [0]{\spacefactor3000\relax}%
\providecommand \BibitemShut  [1]{\csname bibitem#1\endcsname}%
\let\auto@bib@innerbib\@empty
%</preamble>
\bibitem [{\citenamefont {O’Hern}\ \emph {et~al.}(2003)\citenamefont
  {O’Hern}, \citenamefont {Silbert}, \citenamefont {Liu},\ and\ \citenamefont
  {Nagel}}]{ohern_jamming_2003}%
  \BibitemOpen
  \bibfield  {author} {\bibinfo {author} {\bibfnamefont {C.~S.}\ \bibnamefont
  {O’Hern}}, \bibinfo {author} {\bibfnamefont {L.~E.}\ \bibnamefont
  {Silbert}}, \bibinfo {author} {\bibfnamefont {A.~J.}\ \bibnamefont {Liu}}, \
  and\ \bibinfo {author} {\bibfnamefont {S.~R.}\ \bibnamefont {Nagel}},\ }\href
  {\doibase 10.1103/PhysRevE.68.011306} {\bibfield  {journal} {\bibinfo
  {journal} {Physical Review E}\ }\textbf {\bibinfo {volume} {68}},\ \bibinfo
  {pages} {011306} (\bibinfo {year} {2003})}\BibitemShut {NoStop}%
\bibitem [{\citenamefont {Bernal}\ \emph {et~al.}(1962)\citenamefont {Bernal},
  \citenamefont {Mason},\ and\ \citenamefont {Knight}}]{bernal_radial_1962}%
  \BibitemOpen
  \bibfield  {author} {\bibinfo {author} {\bibfnamefont {J.~D.}\ \bibnamefont
  {Bernal}}, \bibinfo {author} {\bibfnamefont {J.}~\bibnamefont {Mason}}, \
  and\ \bibinfo {author} {\bibfnamefont {K.~R.}\ \bibnamefont {Knight}},\
  }\href {\doibase 10.1038/194957a0} {\bibfield  {journal} {\bibinfo  {journal}
  {Nature}\ }\textbf {\bibinfo {volume} {194}},\ \bibinfo {pages} {957}
  (\bibinfo {year} {1962})}\BibitemShut {NoStop}%
\bibitem [{\citenamefont {Donev}\ \emph
  {et~al.}(2005{\natexlab{a}})\citenamefont {Donev}, \citenamefont {Torquato},\
  and\ \citenamefont {Stillinger}}]{donev_pair_2005}%
  \BibitemOpen
  \bibfield  {author} {\bibinfo {author} {\bibfnamefont {A.}~\bibnamefont
  {Donev}}, \bibinfo {author} {\bibfnamefont {S.}~\bibnamefont {Torquato}}, \
  and\ \bibinfo {author} {\bibfnamefont {F.~H.}\ \bibnamefont {Stillinger}},\
  }\href {\doibase 10.1103/PhysRevE.71.011105} {\bibfield  {journal} {\bibinfo
  {journal} {Physical Review E}\ }\textbf {\bibinfo {volume} {71}},\ \bibinfo
  {pages} {011105} (\bibinfo {year} {2005}{\natexlab{a}})}\BibitemShut
  {NoStop}%
\bibitem [{\citenamefont {Morse}\ and\ \citenamefont
  {Corwin}(2014)}]{morse_geometric_2014}%
  \BibitemOpen
  \bibfield  {author} {\bibinfo {author} {\bibfnamefont {P.~K.}\ \bibnamefont
  {Morse}}\ and\ \bibinfo {author} {\bibfnamefont {E.~I.}\ \bibnamefont
  {Corwin}},\ }\href {\doibase 10.1103/PhysRevLett.112.115701} {\bibfield
  {journal} {\bibinfo  {journal} {Physical Review Letters}\ }\textbf {\bibinfo
  {volume} {112}},\ \bibinfo {pages} {115701} (\bibinfo {year}
  {2014})}\BibitemShut {NoStop}%
\bibitem [{\citenamefont {Morse}\ and\ \citenamefont
  {Corwin}(2016)}]{morse_geometric_2016}%
  \BibitemOpen
  \bibfield  {author} {\bibinfo {author} {\bibfnamefont {P.~K.}\ \bibnamefont
  {Morse}}\ and\ \bibinfo {author} {\bibfnamefont {E.~I.}\ \bibnamefont
  {Corwin}},\ }\href {\doibase 10.1039/C5SM02575C} {\bibfield  {journal}
  {\bibinfo  {journal} {Soft Matter}\ }\textbf {\bibinfo {volume} {12}},\
  \bibinfo {pages} {1248} (\bibinfo {year} {2016})}\BibitemShut {NoStop}%
\bibitem [{\citenamefont {Maxwell}(1864)}]{maxwell_calculation_1864}%
  \BibitemOpen
  \bibfield  {author} {\bibinfo {author} {\bibfnamefont {J.~C.}\ \bibnamefont
  {Maxwell}},\ }\href {\doibase 10.1080/14786446408643668} {\bibfield
  {journal} {\bibinfo  {journal} {Philosophical Magazine Series 4}\ }\textbf
  {\bibinfo {volume} {27}},\ \bibinfo {pages} {294} (\bibinfo {year}
  {1864})}\BibitemShut {NoStop}%
\bibitem [{\citenamefont {Donev}\ \emph
  {et~al.}(2005{\natexlab{b}})\citenamefont {Donev}, \citenamefont
  {Stillinger},\ and\ \citenamefont {Torquato}}]{donev_unexpected_2005}%
  \BibitemOpen
  \bibfield  {author} {\bibinfo {author} {\bibfnamefont {A.}~\bibnamefont
  {Donev}}, \bibinfo {author} {\bibfnamefont {F.~H.}\ \bibnamefont
  {Stillinger}}, \ and\ \bibinfo {author} {\bibfnamefont {S.}~\bibnamefont
  {Torquato}},\ }\href {\doibase 10.1103/PhysRevLett.95.090604} {\bibfield
  {journal} {\bibinfo  {journal} {Physical Review Letters}\ }\textbf {\bibinfo
  {volume} {95}},\ \bibinfo {pages} {090604} (\bibinfo {year}
  {2005}{\natexlab{b}})}\BibitemShut {NoStop}%
\bibitem [{\citenamefont {Dreyfus}\ \emph {et~al.}(2015)\citenamefont
  {Dreyfus}, \citenamefont {Xu}, \citenamefont {Still}, \citenamefont {Hough},
  \citenamefont {Yodh},\ and\ \citenamefont
  {Torquato}}]{dreyfus_diagnosing_2015}%
  \BibitemOpen
  \bibfield  {author} {\bibinfo {author} {\bibfnamefont {R.}~\bibnamefont
  {Dreyfus}}, \bibinfo {author} {\bibfnamefont {Y.}~\bibnamefont {Xu}},
  \bibinfo {author} {\bibfnamefont {T.}~\bibnamefont {Still}}, \bibinfo
  {author} {\bibfnamefont {L.~A.}\ \bibnamefont {Hough}}, \bibinfo {author}
  {\bibfnamefont {A.~G.}\ \bibnamefont {Yodh}}, \ and\ \bibinfo {author}
  {\bibfnamefont {S.}~\bibnamefont {Torquato}},\ }\href {\doibase
  10.1103/PhysRevE.91.012302} {\bibfield  {journal} {\bibinfo  {journal}
  {Physical Review E}\ }\textbf {\bibinfo {volume} {91}},\ \bibinfo {pages}
  {012302} (\bibinfo {year} {2015})}\BibitemShut {NoStop}%
\bibitem [{\citenamefont {Clusel}\ \emph {et~al.}(2009)\citenamefont {Clusel},
  \citenamefont {Corwin}, \citenamefont {Siemens},\ and\ \citenamefont
  {Brujić}}]{clusel_granocentric_2009}%
  \BibitemOpen
  \bibfield  {author} {\bibinfo {author} {\bibfnamefont {M.}~\bibnamefont
  {Clusel}}, \bibinfo {author} {\bibfnamefont {E.~I.}\ \bibnamefont {Corwin}},
  \bibinfo {author} {\bibfnamefont {A.~O.~N.}\ \bibnamefont {Siemens}}, \ and\
  \bibinfo {author} {\bibfnamefont {J.}~\bibnamefont {Brujić}},\ }\href
  {\doibase 10.1038/nature08158} {\bibfield  {journal} {\bibinfo  {journal}
  {Nature}\ }\textbf {\bibinfo {volume} {460}},\ \bibinfo {pages} {611}
  (\bibinfo {year} {2009})}\BibitemShut {NoStop}%
\bibitem [{\citenamefont {Maiti}\ and\ \citenamefont
  {Sastry}(2014)}]{maiti_free_2014}%
  \BibitemOpen
  \bibfield  {author} {\bibinfo {author} {\bibfnamefont {M.}~\bibnamefont
  {Maiti}}\ and\ \bibinfo {author} {\bibfnamefont {S.}~\bibnamefont {Sastry}},\
  }\href {\doibase 10.1063/1.4891358} {\bibfield  {journal} {\bibinfo
  {journal} {The Journal of Chemical Physics}\ }\textbf {\bibinfo {volume}
  {141}},\ \bibinfo {pages} {044510} (\bibinfo {year} {2014})}\BibitemShut
  {NoStop}%
\bibitem [{\citenamefont {Parisi}\ and\ \citenamefont
  {Zamponi}(2010)}]{parisi_mean-field_2010}%
  \BibitemOpen
  \bibfield  {author} {\bibinfo {author} {\bibfnamefont {G.}~\bibnamefont
  {Parisi}}\ and\ \bibinfo {author} {\bibfnamefont {F.}~\bibnamefont
  {Zamponi}},\ }\href {\doibase 10.1103/RevModPhys.82.789} {\bibfield
  {journal} {\bibinfo  {journal} {Reviews of Modern Physics}\ }\textbf
  {\bibinfo {volume} {82}},\ \bibinfo {pages} {789} (\bibinfo {year}
  {2010})}\BibitemShut {NoStop}%
\bibitem [{\citenamefont {Charbonneau}\ \emph {et~al.}(2014)\citenamefont
  {Charbonneau}, \citenamefont {Kurchan}, \citenamefont {Parisi}, \citenamefont
  {Urbani},\ and\ \citenamefont {Zamponi}}]{charbonneau_fractal_2014}%
  \BibitemOpen
  \bibfield  {author} {\bibinfo {author} {\bibfnamefont {P.}~\bibnamefont
  {Charbonneau}}, \bibinfo {author} {\bibfnamefont {J.}~\bibnamefont
  {Kurchan}}, \bibinfo {author} {\bibfnamefont {G.}~\bibnamefont {Parisi}},
  \bibinfo {author} {\bibfnamefont {P.}~\bibnamefont {Urbani}}, \ and\ \bibinfo
  {author} {\bibfnamefont {F.}~\bibnamefont {Zamponi}},\ }\href {\doibase
  10.1038/ncomms4725} {\bibfield  {journal} {\bibinfo  {journal} {Nature
  Communications}\ }\textbf {\bibinfo {volume} {5}} (\bibinfo {year} {2014}),\
  10.1038/ncomms4725}\BibitemShut {NoStop}%
\bibitem [{\citenamefont {O'Hern}\ \emph {et~al.}(2002)\citenamefont {O'Hern},
  \citenamefont {Langer}, \citenamefont {Liu},\ and\ \citenamefont
  {Nagel}}]{ohern_random_2002}%
  \BibitemOpen
  \bibfield  {author} {\bibinfo {author} {\bibfnamefont {C.~S.}\ \bibnamefont
  {O'Hern}}, \bibinfo {author} {\bibfnamefont {S.~A.}\ \bibnamefont {Langer}},
  \bibinfo {author} {\bibfnamefont {A.~J.}\ \bibnamefont {Liu}}, \ and\
  \bibinfo {author} {\bibfnamefont {S.~R.}\ \bibnamefont {Nagel}},\ }\href
  {\doibase 10.1103/PhysRevLett.88.075507} {\bibfield  {journal} {\bibinfo
  {journal} {Physical Review Letters}\ }\textbf {\bibinfo {volume} {88}},\
  \bibinfo {pages} {075507} (\bibinfo {year} {2002})}\BibitemShut {NoStop}%
\bibitem [{\citenamefont {Hestenes}\ and\ \citenamefont
  {Stiefel}(1952)}]{hestenes_methods_1952}%
  \BibitemOpen
  \bibfield  {author} {\bibinfo {author} {\bibfnamefont {M.~R.}\ \bibnamefont
  {Hestenes}}\ and\ \bibinfo {author} {\bibfnamefont {E.}~\bibnamefont
  {Stiefel}},\ }\href@noop {} {\bibfield  {journal} {\bibinfo  {journal}
  {Journal Of Research Of The National Bureau Of Standards}\ }\textbf {\bibinfo
  {volume} {49}},\ \bibinfo {pages} {409} (\bibinfo {year} {1952})}\BibitemShut
  {NoStop}%
\bibitem [{\citenamefont {Bitzek}\ \emph {et~al.}(2006)\citenamefont {Bitzek},
  \citenamefont {Koskinen}, \citenamefont {Gähler}, \citenamefont {Moseler},\
  and\ \citenamefont {Gumbsch}}]{bitzek_structural_2006}%
  \BibitemOpen
  \bibfield  {author} {\bibinfo {author} {\bibfnamefont {E.}~\bibnamefont
  {Bitzek}}, \bibinfo {author} {\bibfnamefont {P.}~\bibnamefont {Koskinen}},
  \bibinfo {author} {\bibfnamefont {F.}~\bibnamefont {Gähler}}, \bibinfo
  {author} {\bibfnamefont {M.}~\bibnamefont {Moseler}}, \ and\ \bibinfo
  {author} {\bibfnamefont {P.}~\bibnamefont {Gumbsch}},\ }\href {\doibase
  10.1103/PhysRevLett.97.170201} {\bibfield  {journal} {\bibinfo  {journal}
  {Physical Review Letters}\ }\textbf {\bibinfo {volume} {97}},\ \bibinfo
  {pages} {170201} (\bibinfo {year} {2006})}\BibitemShut {NoStop}%
\bibitem [{\citenamefont {Boyd}\ and\ \citenamefont
  {Vandenberghe}(2004)}]{boyd_convex_2004}%
  \BibitemOpen
  \bibfield  {author} {\bibinfo {author} {\bibfnamefont {S.}~\bibnamefont
  {Boyd}}\ and\ \bibinfo {author} {\bibfnamefont {L.}~\bibnamefont
  {Vandenberghe}},\ }\href@noop {} {\emph {\bibinfo {title} {Convex
  {Optimization}}}}\ (\bibinfo  {publisher} {Cambridge University Press},\
  \bibinfo {year} {2004})\BibitemShut {NoStop}%
\bibitem [{\citenamefont {Boissonnat}\ and\ \citenamefont
  {Delage}(2005)}]{boissonnat_convex_2005}%
  \BibitemOpen
  \bibfield  {author} {\bibinfo {author} {\bibfnamefont {J.-D.}\ \bibnamefont
  {Boissonnat}}\ and\ \bibinfo {author} {\bibfnamefont {C.}~\bibnamefont
  {Delage}},\ }in\ \href
  {http://www.springerlink.com/index/10.1007/11561071_34} {\emph {\bibinfo
  {booktitle} {Algorithms – {ESA} 2005}}},\ Vol.\ \bibinfo {volume} {3669}\
  (\bibinfo  {publisher} {Springer Berlin Heidelberg},\ \bibinfo {address}
  {Berlin, Heidelberg},\ \bibinfo {year} {2005})\ pp.\ \bibinfo {pages}
  {367--378}\BibitemShut {NoStop}%
\bibitem [{\citenamefont {Voronoi}(1908)}]{voronoi_nouvelles_1908}%
  \BibitemOpen
  \bibfield  {author} {\bibinfo {author} {\bibfnamefont {G.}~\bibnamefont
  {Voronoi}},\ }\href
  {http://www.degruyter.com/view/j/crll.1908.issue-134/crll.1908.134.198/crll.1908.134.198.xml}
  {\bibfield  {journal} {\bibinfo  {journal} {Journal Fur Die Reine Und
  Angewandte Mathematik}\ ,\ \bibinfo {pages} {198}} (\bibinfo {year}
  {1908})}\BibitemShut {NoStop}%
\bibitem [{\citenamefont {Charbonneau}\ \emph {et~al.}(2012)\citenamefont
  {Charbonneau}, \citenamefont {Corwin}, \citenamefont {Parisi},\ and\
  \citenamefont {Zamponi}}]{charbonneau_universal_2012}%
  \BibitemOpen
  \bibfield  {author} {\bibinfo {author} {\bibfnamefont {P.}~\bibnamefont
  {Charbonneau}}, \bibinfo {author} {\bibfnamefont {E.~I.}\ \bibnamefont
  {Corwin}}, \bibinfo {author} {\bibfnamefont {G.}~\bibnamefont {Parisi}}, \
  and\ \bibinfo {author} {\bibfnamefont {F.}~\bibnamefont {Zamponi}},\ }\href
  {\doibase 10.1103/PhysRevLett.109.205501} {\bibfield  {journal} {\bibinfo
  {journal} {Physical Review Letters}\ }\textbf {\bibinfo {volume} {109}},\
  \bibinfo {pages} {205501} (\bibinfo {year} {2012})}\BibitemShut {NoStop}%
\bibitem [{\citenamefont {Onoda}\ and\ \citenamefont
  {Liniger}(1990)}]{onoda_random_1990}%
  \BibitemOpen
  \bibfield  {author} {\bibinfo {author} {\bibfnamefont {G.~Y.}\ \bibnamefont
  {Onoda}}\ and\ \bibinfo {author} {\bibfnamefont {E.~G.}\ \bibnamefont
  {Liniger}},\ }\href {\doibase 10.1103/PhysRevLett.64.2727} {\bibfield
  {journal} {\bibinfo  {journal} {Physical Review Letters}\ }\textbf {\bibinfo
  {volume} {64}},\ \bibinfo {pages} {2727} (\bibinfo {year}
  {1990})}\BibitemShut {NoStop}%
\bibitem [{\citenamefont {Song}\ \emph {et~al.}(2008)\citenamefont {Song},
  \citenamefont {Wang},\ and\ \citenamefont {Makse}}]{song_phase_2008}%
  \BibitemOpen
  \bibfield  {author} {\bibinfo {author} {\bibfnamefont {C.}~\bibnamefont
  {Song}}, \bibinfo {author} {\bibfnamefont {P.}~\bibnamefont {Wang}}, \ and\
  \bibinfo {author} {\bibfnamefont {H.~A.}\ \bibnamefont {Makse}},\ }\href
  {\doibase 10.1038/nature06981} {\bibfield  {journal} {\bibinfo  {journal}
  {Nature}\ }\textbf {\bibinfo {volume} {453}},\ \bibinfo {pages} {629}
  (\bibinfo {year} {2008})}\BibitemShut {NoStop}%
\bibitem [{\citenamefont {Parisi}\ and\ \citenamefont
  {Zamponi}(2005)}]{parisi_ideal_2005}%
  \BibitemOpen
  \bibfield  {author} {\bibinfo {author} {\bibfnamefont {G.}~\bibnamefont
  {Parisi}}\ and\ \bibinfo {author} {\bibfnamefont {F.}~\bibnamefont
  {Zamponi}},\ }\href {\doibase doi:10.1063/1.2041507} {\bibfield  {journal}
  {\bibinfo  {journal} {The Journal of Chemical Physics}\ }\textbf {\bibinfo
  {volume} {123}},\ \bibinfo {pages} {144501} (\bibinfo {year}
  {2005})}\BibitemShut {NoStop}%
\end{thebibliography}%

\end{document}